\documentclass{elsart}
\usepackage{graphicx,amssymb}
\usepackage{epsfig}
\usepackage{graphicx}

\journal{Physica A}

\begin{document}

\begin{frontmatter}
\title{Protein motors induced enhanced diffusion in intracellular transport}

\author{I. Santamar\'{i}a-Holek,$^a$ M. H. Vainstein,$^b$
J. M. Rub\'{\i}$^c$ and F. A. Oliveira$^{b}$}
\address{$^a$ Facultad de Ciencias, Universidad Nacional Aut\'{o}noma de
M\'{e}xico.\\
Circuito exterior de Ciudad Universitaria. 04510, D. F.,
M\'{e}xico.}
\address{$^b$ Institute of Physics and International Center of Condensed Matter\\
Physics University of Bras\'{\i}lia, CP 04513, 70919-970,
Bras\'{\i}lia-DF, Brazil.}
\address{$^c$ Facultat de F\'{\i}sica, Universitat de Barcelona. \\
Av. Diagonal 647, 08028, Barcelona, Spain.  }

\thanks[fn1]{E-mail: isholek.fc@gmail.com}

\begin{abstract} Diffusion of transported particles
in the intracellular medium is described by means of a generalized
diffusion equation containing forces due to the cytoskeleton
network and to the protein motors. We find that the enhanced
diffusion observed in experiments depends on the nature of the
force exerted by the protein motors and on parameters
characterizing the intracellular medium which is described in
terms of a generalized Debye spectrum for the noise density of
states.
\end{abstract}
\begin{keyword}
Fokker-Planck equations; enhanced diffusion, molecular motors.
\end{keyword}
\end{frontmatter}

\section{Introduction}

\label{intro}

The study of the spatio-temporal organization of vesicles,
organelles and other particles inside the cellular medium plays a
primary role in the behaviour of cells and constitutes a central
problem in physical biology \cite{vale,Fear}.

In order to carry out this study, it is important to consider the
fact that the intracellular medium consists of a wide variety of
polymers and particles whose presence is the cause of the observed
viscolestic behaviour \cite{lubensky}. As a consequence, during
their motion inside the cell the particles may experience elastic,
electrochemical and entropic forces that introduce memory effects.
These effects must be taken into account in order to correctly
describe the dynamics of the particles.

Particles in the intracellular medium undergo different forms of
diffusion which depends on the activity of the protein motors and
the size of the particles
\cite{caspi,thermokinetic,adelman,oxtoby,Chaudhury,finite,WEITZNATURE06,wong,
blockNature, rev-reimann,oliveira,oliveira2,Adam07}. Subdiffusion
has been observed in the passive transport of particles in the
cellular medium
\cite{caspi,thermokinetic,finite,WEITZNATURE06,wong} whereas an
enhanced diffusion has been reported for particles driven by
protein motors \cite{lubensky,caspi,blockNature}.

A complete description of particle transport must take into
account the activity of protein motors
\cite{lubensky,caspi,blockNature}. These entities are
chemo-mechanical transductors which, in the presence of
sufficiently high cation concentration, trap particles and then
use the free energy arising from the hydrolysis of ATP molecules
to drive them along defined trajectories inside the cell
\cite{vale,demiguel1,peterman} giving rise to the enhanced
diffusion observed in these systems \cite{lubensky,caspi}.
Experiments seem to justify the assumption that the activation as
well as the duration of the activity of these protein motors
constitutes a random process \cite{lubensky}.

In this article, we use mesoscopic nonequilibrium thermodynamics
\cite{REVIEWjpcb,pnas,groot} to formulate a mesoscopic model that
accounts for the  enhanced diffusion of particles in the
intracellular medium. We propose a generalized diffusion equation,
that incorporates the memory effects due to the viscoelasticity of
the intracellular medium and the activity of protein motors, to
compute the mean square displacement (MSD) and to analyze the
density of states of the intracellular medium
\cite{thermokinetic,finite,balescu}.  Memory effects can be
incorporated through \emph{effective} mobility and diffusion
coefficients that can be memory kernels,
\cite{thermokinetic,balescu,z-libro,rev-fractional} or
time-dependent coefficients
\cite{adelman,oxtoby,Chaudhury,finite,balescu}. In our approach,
the generalized diffusion equation incorporates memory effects
through time dependent coefficients. We choose this way because in
the linear force case it is possible to show the equivalence of
this description with that given by means of generalized Langevin
equations with memory kernels \cite{adelman,oxtoby}. This relation
between both descriptions is also used in order to calculate the
noise density of states arising in the case of enhanced diffusion.
However, in this case the stochastic forces are of non-thermal
nature due to the activity of protein motors. It is convenient to
mention here that, in the general case, there is not a clear
criterion in order to choose between time-dependent coefficients
and memory kernels, see, for example, Ref. \cite{marcus}.

The article is organized as follows. In Sec. \textbf{2}, we derive a
generalized diffusion equation (GDE) describing the
dynamics of the system. In section \textbf{3}, we analyze the
case of a harmonic force model and compare the results
obtained with experiments. Section \textbf{4} is devoted to derive
the explicit form of the noise density of states (NDS). In Sec.
\textbf{5} we summarize and discuss our main results.

\section{Mesoscopic thermodynamics approach to anomalous diffusion}

\label{MNET}

We consider a Brownian particle moving in an intracellular medium
having viscoelastic nature \cite{lubensky}. In addition, we will
assume that under the appropriate conditions the particle can also
be driven by protein motors. As we have mentioned previously,
these factors introduce memory effects that will be taken into
account through the \emph{effective} time dependent mobility and
diffusion coefficients.

These effects can appropriately be described in terms of a
generalized diffusion equation for the single particle
distribution function $f(\vec{r},t)$, which gives the probability
of finding the particle at position $\vec{r}$ at time $t$. This
distribution function satisfies the normalization condition $\int
f(\vec{r},t)d\vec{r}=1$,  and its evolution in time is governed by
the continuity equation
\begin{equation}
\frac{\partial }{\partial t}f(\vec{r},t)=-\nabla\cdot [f(\vec{r},t) \vec{V}_{r}(\vec{r},t)],
\label{continuidad}
\end{equation}
which expresses probability conservation and where
$\nabla=\partial/\partial \vec{r}$. The quantity $f \vec{V}_{r}$
represents an unknown probability current and the explicit
expression of the streaming velocity $\vec{V}_{r}$  can be
calculated with mesoscopic nonequilibrium thermodynamics (MNET)
using the generalized Gibbs entropy postulate
\cite{thermokinetic,REVIEWjpcb,pnas}
\begin{equation}
S(t)=-k_{B}\int f\ln \frac{f(\vec{r},t)}{f_0(\vec{r})}d\vec{r}+S_{0},
\label{post. gibbs}
\end{equation}
where $k_{B} $ is Boltzmann's constant and $f_{0}(\vec{r})$ and
$S_{0}$ are the distribution function and the entropy of the
reference local equilibrium state. The local equilibrium
distribution function is given by
$f_{0}(\vec{r})=e^{(\mu_0-\phi)(m/k_BT)}$, in which $m$ is mass of
the particle, $T$ is the temperature of the system,
$\phi(\vec{r})$ is an external potential per unit mass and $\mu_0$
is the chemical potential per unit mass at local equilibrium .

Taking the time derivative of Eq. (\ref{post. gibbs})
and using Eq. (\ref{continuidad}) one obtains the expression
$\partial S(t)/\partial t= -k_B\int \ln \left[{f}/{f_0}\right]
 \nabla\cdot (f \vec{V}_{r}) d\vec{r} $. Now, integrating by parts
assuming that the current vanishes at the boundary, we find the
entropy production $\sigma (t)$
\begin{equation}
\sigma(t) =-\frac{m}{T}\int f(\vec{r},t)
\vec{V}_{r}(\vec{r},t)\cdot \nabla \mu(\vec{r},t) d\vec{r}\geq0,
\label{sigma}
\end{equation}
where $\mu(\vec{r},t)=({k_{B}T}/{m}) \ln f/f_{0}+\mu_{0}$ is the
nonequilibrium chemical potential per unit mass. According to
nonequilibrium thermodynamics, we will assume that the force
$\nabla \mu(\vec{r},t)$ and the current $f(\vec{r},t)
\vec{V}_{r}(\vec{r},t)$ are coupled linearly: $f \vec{V}_{r}
\propto \nabla \mu$, \cite{thermokinetic,finite,pnas,groot}. Using
the expression for $f_0(\vec{r})$ one obtains
\begin{equation}
f(\vec{r},t) \vec{V}_{r}(\vec{r},t)= \beta^{-1}(t)f(\vec{r},t)
\left(\vec{F}(\vec{r}) - \frac{k_{B}T}{m}  \nabla \ln f(\vec{r},t)
\right), \label{liner-rel}
\end{equation}
where $\vec{F}(\vec{r})=-\nabla \phi(\vec{r})$ is the total force
applied on the particle.  As mentioned previously, the time
dependent Onsager coefficient $\beta^{-1}(t)$ accounts for the
presence of memory effects. In general, this quantity is related
with the correlation function of the position vector of the
particle. We have chosen this coupling between the current and
force because in the linear force case our results are equivalent
to those obtained from a GLE in Refs. [7,8]. The case when the
coupling is given through a memory kernel has been discussed in
Ref. [6]. In the absence of memory effects, $\beta^{-1}(t)$
reduces to the Stokes mobility  $\beta_0^{-1} = (6 \pi a
\eta_s/m)^{-1}$ with $a$ and $m$ the radius and mass of the
particle and $\eta_s$ the viscosity of the medium.
\cite{adelman,hynes}   Substituting Eq. (\ref{liner-rel}) into
(\ref{continuidad}), we finally obtain the generalized diffusion
equation
\begin{equation}
\frac{\partial }{\partial t}f(\vec{r},t)= \frac{k_{B}T}{m}
\beta^{-1}(t) \nabla^2 f(\vec{r},t) - \beta^{-1}(t) \nabla
\cdot\left[f(\vec{r},t) \vec{F}(\vec{r}) \right]. \label{G-Smol}
\end{equation}
A similar equation has been proposed to describe finite-size and
confinement effects on the dynamics of passively diffusing
particles moving through the intracellular medium.
\cite{thermokinetic,finite}

\section{Enhanced diffusion in the intracellular medium}
\label{Sec2}

A physical model describing the dynamics
of a particle driven by protein
motors must take into account two forces. The first one
is a trapping force of elastic nature which in first
approximation can be modelled by the linear force:
$\vec{F}_{el}(\vec{r})=-\omega_0^2 \vec{r}$, with $\omega_0$ a
characteristic frequency. The second one is the driving force
exerted by molecular motors on the particles that can also
be modelled by
$\vec{F}_{pm}(\vec{r})=\omega_1^2 \vec{r}$, with a
characteristic frequency $\omega_1$.

Substitution of the total force
$F(\vec{r})=\vec{F}_{el}(\vec{r})+\vec{F}_{pm}(\vec{r})$
into Eq. (\ref{G-Smol}) yields
\begin{eqnarray}
\frac{\partial }{\partial t}f(\vec{r},t)= \beta^{-1}(t)
\left[(\omega_0^2-\omega_1^2) \nabla \cdot \left[\vec{r}
f(\vec{r},t)\right] + \frac{k_{B}T}{m} \nabla^2 f(\vec{r},t)
\right] , \label{G-Smol-Osc}
\end{eqnarray}
where we have assumed that the effective friction coefficient is
given by $\beta(t)=\beta_0 \tilde{\beta}(t)$.
$\tilde{\beta}(t)$ is a dimensionless function accounting for
memory effects. Eq. (\ref{G-Smol-Osc}) clearly manifests the
competition between trapping and driving forces that will regulate
the behavior of the MSD of the particle. By introducing the
dimensionless variables $\vec{x}=a^{-1}\vec{r}$ and
$\tilde{t}=\tau^{-1}_D t$, with $\tau_D$ a characteristic time, and
scaling time with $z(t)=\int^t_0 \tilde{\beta}^{-1}(t')dt'$,
we obtain
\begin{eqnarray}
\langle x^2(\tilde{t})\rangle = 3\frac{\omega_T^2}{(\omega_0^2-\omega_1^2)}
\left[1- e^{-2(\omega_0^2-\omega_1^2)\frac{\tau_D}{\beta_0}z(\tilde{t})
}\right] , \label{MSD}
\end{eqnarray}
where we have introduced the constant $\omega_T^2=k_BT/ma^2$ and
$\langle x^2(\tilde{t})\rangle=\int x^2 f d\vec{x}$. Eq.
(\ref{MSD}) gives the behavior of the MSD in  the case when
$\omega_0^2-\omega_1^2>0$, that is, when the friction is larger
than the force exerted by the protein motors. The case when
$\omega_0^2-\omega_1^2<0$ will be described later.

The time dependence of $z(t)$ is related to the behavior of the
time correlation function $\chi(\tilde{t})=\langle
\vec{x}\cdot\vec{x}_0\rangle(\tilde{t})$ in the absence of
molecular motors, in the subdiffusion regime $\omega_1=0$. The
evolution equation for $\chi(\tilde{t})$ is given by
\cite{adelman,oxtoby,finite}
\begin{equation}
\frac{d }{d \tilde{t}} \chi(\tilde{t}) = -
\,\omega_0^2\frac{\tau_D}{\beta_0}
\tilde{\beta}^{-1}(t)\chi(\tilde{t}), \label{evol-Xi}
\end{equation}
which can also be calculated by using the dimensionless form of
Eq. (\ref{G-Smol-Osc}). Now by taking into account the expression
of $z(t)$ in terms of $\tilde{\beta}^{-1}$, after solving
(\ref{evol-Xi}) one arrives at the relation
\begin{equation}
 z_{sub}(\tilde{t}) =
-\,\frac{\beta_0}{\tau_D\omega_0^2} \ln R(\tilde{t}),
\label{tau-Xi1}
\end{equation}
where we have defined the normalized correlation
$R(\tilde{t})=\chi(\tilde{t})/\chi(0)$. From this relation it
follows that the scaling of
time depends on the relaxation dynamics of the system.

At short times, one may obtain the behavior of
$z(\tilde{t})$ by expanding the logarithm in its argument around
one: $\ln| R^{-\,{\beta_0}/{\tau_D\omega_0^2}}|\simeq
R^{-{\beta_0}/{\tau_D\omega_0^2}}-1 + O(R^2)$. Then taking into
account that $R(\tilde{t})$ must be an even function of time
 \cite{oliveira2}, a first order expansion leads to:
$R^{-1}(\tilde{t})\simeq 1+B_2^*\tilde{t}^2+O(\tilde{t}^4)$. Thus,
at short times we obtain the expression
\begin{equation}
z_{sub}(\tilde{t}) \sim B_2 \tilde{t}^{\,{2\beta_0}/{\tau_D\omega_0^2}},
\label{tau-Xi}
\end{equation}
 where $ B_2 \propto B_2^*$ is a parameter characterizing the type of
relaxation \cite{finite}. Now, by expanding the exponential in
Eq.(\ref{MSD}) up to first order in its argument and substituting
(\ref{tau-Xi}) into the result, we find
\begin{equation}
\langle x^2(\tilde{t})\rangle_{sub} \simeq 6B_2 \tau_D
\frac{\omega_T^2}{\beta_0}
\tilde{t}^{\,{2\beta_0}/{\tau_D\omega_0^2}} .\label{MSD2}
\end{equation}
The exponent characterizing the subdiffusion process of the
particle is a function of parameters of the bath and the forces
acting on the particle \cite{wong,finite}. The coefficient $B_2$
depends in general on the concentration of polymers in the medium
and the size of the particles, thus characterizing the magnitude
of the MSD, \cite{finite}. For $t=\tau_D$, Eq. (\ref{MSD2}) yields
$B_2=(\beta_0/\tau_D\omega_0^2)\langle x^2(1)\rangle_{sub}$. Using
the definition of the running diffusion coefficient
$D_{sub}(\tilde{t})=d\langle x^2\rangle(\tilde{t})/d\tilde{t}$,
\cite{balescu}, Eq. (\ref{MSD}) leads to the relation
\begin{equation}
\tilde{\beta}_{sub}^{-1}(\tilde{t}) \simeq
\tilde{t}^{-1+{\,{2\beta_0}/{\tau_D\omega_0^2}}}.
\label{beta-tilde}
\end{equation}
This expression can be used to obtain the shear modulus
characterizing the viscoelastic properties of the material
\cite{thermokinetic,finite}.

In Fig. {\bf 1}, we compare our results with experiments (symbols)
reported in Ref. \cite{caspi}, where the motion of a microsphere
through a living eukaryotic cell was observed with video-based
methods. The value of $\omega_0$ can be determined by considering
the case when the particle performs subdiffusion or is attached to
the cytoskeleton ($\omega_1=0$). From the experimental results
represented by open circles in Fig. {\bf 1} it follows that the
short time behavior of the MSD satisfies a power law with exponent
$\sim 3/4$, \cite{caspi,WEITZNATURE06,libroAndelman}. Using the
parameters $B_2$ and $\tau_D$ to fit the curve (dot-dashed line)
with Eqs. (\ref{MSD}) and (\ref{tau-Xi}), one obtains
$\omega_0^2\sim (8/3)\beta_0\tau_D^{-1}$, implying that $\omega_0$
is proportional to the geometric mean of the fast $\sim \beta_0$
and slow $\sim \tau_D^{-1}$ modes (see the caption of Fig. {\bf
1}). Since $\omega_0$ describes subdiffusion, it must satisfy the
condition $\omega_0\geq\sqrt{2\beta_0/\tau_D}$. It is worth
noticing that the exponent depends on the friction coefficient
$\beta_0$, the characteristic frequency $\omega_0$ of the elastic
force exerted by the medium on the particle and a scaling time
$\tau_D$ which is related with the time at which the MSD
saturates. In a previous study this power law behavior was
obtained by simply inferring the exponent from direct comparison
with the experimental results \cite{caspi}. Here, we show that it
is connected with the interactions between the particles and the
medium.

The observed enhancement of diffusion can be described by taking
into account the activity of the motors ($\omega_1^2\neq0$) in Eq. (\ref{G-Smol-Osc}). Defining $\tilde{\omega}^2=\omega_1^2-\omega_0^2>0$, we use
Eq. (\ref{G-Smol-Osc}) in order to obtain a linear evolution equation
for $\langle\vec{x}(\tilde{t})\rangle=\int\vec{x}f(\vec{x},t)d\vec{x}$ whose
solution is $\langle\vec{x}(\tilde{t})\rangle=
exp\left[\tilde{\omega}^2\tau_D\beta_0^{-1}\int\tilde{\beta}_{sub}^{-1}(\tilde{t}')d\tilde{t}'\right]$. Assuming that $\tilde{\omega}^2\tau_D\beta_0^{-1}<1$, an expansion of the exponential up to first order yields
\begin{equation}
\langle\vec{x}(\tilde{t})\rangle \simeq
1+\tilde{\omega}^2\tau_D\beta_0^{-1}\langle x^2(\tilde{t})\rangle_{sub},
\label{x-prom}
\end{equation}
where we have used the relation
$z_{sub}(\tilde{t})=\int\tilde{\beta}_{sub}^{-1}(\tilde{t})d\tilde{t}$.
Considering that, in general, protein motors are randomly
distributed in the cell, we will assume that during the motion of
the particle the parameter $\tilde{\omega}^2$ is a fluctuating
quantity with zero mean and
$\langle\tilde{\omega}^2(\tilde{t})\tilde{\omega}^2(\tilde{t}')\rangle=
\zeta exp\left[-(\tilde{t}-\tilde{t}')/\tau_c\right]$, with
$\tau_c$ the correlation time of the noise and
$\zeta=\omega_1^2-\omega_0^2$ its magnitude. Now, by evaluating
$\langle\vec{x}(\tilde{t})\rangle\langle\vec{x}(\tilde{t}')\rangle$
and taking the average over the realizations of the random force,
represented by $\overline{\langle\vec{x}(\tilde{t})\rangle^2}$,
one finally obtains
\begin{equation}
\overline{\langle\vec{x}(\tilde{t})\rangle^2}\simeq
1+\zeta^2\tau_D^2\beta_0^{-2}\,\tilde{t}^{4\beta_0/\omega_{0}^2\tau_D}.
\label{xx}
\end{equation}
Fig. \textbf{1} shows a comparison between experiments (filled
symbols) and theory (solid and dashed lines) of enhanced diffusion
due to the presence of protein motors. For
$2\beta_0/(\omega_{0}^2\tau_D)=3/4$, from  Eq. (\ref{xx}) it
follows that the exponent characterizing the enhanced diffusion is
$\alpha_{ed} \sim 3/2$. The results (\ref{MSD2}) and
(\ref{xx}) constitute an alternative way to calculate the Hurst
exponent of the MSD typically arising in systems in which
self-avoiding random walks (SAWR) appear frequently, such as
crystallization or cluster-cluster aggregation in a viscoelastic
medium \cite{thermokinetic,Adam-Rev,Ad-TechLetters,Jacek}. This
fact allows one to interpret the present results within the scope
of SAWR formalism \cite{Adam-Rev} in which the exponent obtained
is $3/2=2/d$ with $d$ the dimension of the space. The value of $d$
corresponding to our problem in the enhanced diffusion case is
therefore $d=4/3$. Thus, the molecular motors act as elastic
scatters inducing a SARW in a space with slightly reduced
(motion-oriented) degrees of freedom.

\section{The noise density of states of the intracellular medium}
\label{SecNDS}

A global characterization of the intracellular medium
can be performed through the noise density if states (NDS).
This quantity contains in average form the statistical properties
of the intracellular medium, assimilated to an effective medium in
which both thermal and non-thermal (motor-induced) fluctuations
control the dynamics of the transported particles.

The NDS can be analyzed by means of a generalized Langevin
equation \cite{oliveira,oliveira2} which is equivalent to the following generalized diffusion equation \cite{adelman,Chaudhury}
\begin{eqnarray}
\frac{\partial }{\partial t}f(\vec{r},t)=D_{eff}(t) \nabla^2 f(\vec{r},t),
\label{G-super}
\end{eqnarray}
where $D_{eff}(t)=d\overline{\langle\vec{x}(t)\rangle^2}/dt$ is
an effective diffusion coefficient incorporating the effects of the
motors in average form. Eq. (\ref{G-super}) is equivalent to
the generalized Langevin equation for the velocity $v(\tilde{t})$
of the particle in the overdamped case \cite{adelman,Chaudhury}
\begin{equation}
\int^t_0
\Pi(\tilde{t}-\tilde{t}')v(\tilde{t}')d\tilde{t}' =
F^R(\tilde{t}), \label{GLE}
\end{equation}
where we have considered the one-dimensional case for simplicity
and $F^R$ is a stochastic force satisfying the conditions
$\langle F^R \rangle =0$, $\langle F^R \,x(0)\rangle=0$ and
\begin{equation}
\langle F^R(\tilde{t}) F^R(\tilde{t}')\rangle
= \langle v^2\rangle_{eq} \Pi(\tilde{t}-\tilde{t}'),
\label{FDT}
\end{equation}
with $\langle v^2\rangle_{eq}=\lim_{t\rightarrow\infty}\langle
v^2\rangle(t)$. By Laplace transforming Eqs. (\ref{G-super}) and
(\ref{GLE}) we can obtain the relation between the diffusion
coefficient and the memory function in the following form
\begin{equation}
\label{Ded}
D_{eff}(t)=\frac{k_BT}{m}{\cal L}^{-1}\left[\frac{1}{s\hat{\Pi}(s)}\right],
\end{equation}
where $\hat{\Pi}(s)$ is the Laplace transform of $\Pi(t)$ and
${\cal L}^{-1}[\hat{\Pi}(s)]$ the inverse transform.
Using now the definition of $D_{eff}(t)$ in terms of Eqs. (\ref{xx}) one obtains
\begin{equation}
\label{Pi(t)}
\Pi(t) \sim \,\tilde{t}^{4\beta_0/\omega_{0}^2\tau_D}.
\end{equation}
This equation can be used to calculate the NDS of states
$\rho(\omega)$ of the intracellular medium by decomposing the
stochastic force $F^R(t)$ of the bath in a set of harmonic
oscillators \cite{oliveira,oliveira2}.
The relation between the memory function
and the NDS is found through the fluctuation dissipation theorem. One obtains
\begin{equation}
\Pi(\tilde{t}) = \int \rho(\omega) cos(\omega t)d\omega,
\label{Pi-NDS}
\end{equation}
where $\omega$ is now a dimensionless frequency.

Taking the Fourier cosine transform of Eq. (\ref{Pi-NDS}) with
(\ref{Pi(t)}), the resulting
expression for the NDS of the intracellular medium is
\begin{equation}
\rho(\omega) \sim
\frac{2}{\pi}\,
\Gamma\left[1+\frac{4\beta_0}{\omega_{0}^2\tau_D}\right]\sin \left[\frac{2 \pi \beta_0}{\omega_{0}^2\tau_D}\right]|\omega|^{-1+4\beta_0/\omega_{0}^2\tau_D}.
\label{NDS}
\end{equation}
From this equation it follows that the exponent controlling the
behavior of the NDS depends on the frequency characterizing the
elastic forces ($\omega_0$) in the intracellular medium. It is
important to stress that the power law dependence on the frequency
of the NDS given by (\ref{NDS}) and valid up to the corresponding
Debye cutoff frequency, is of the form of the generalized Debye
spectrum proposed in Ref. \cite{oliveira2}.

The results we have obtained enable us to conclude that the
activity of protein motors in the intracellular medium modifies
the dependence on frequency of the NDS. For example, in accordance
with experiments \cite{lubensky,caspi} the MSD of the particles
grows by following the power law
$\overline{\langle\vec{x}(\tilde{t})\rangle^2} \sim
\tilde{t}^{3/2}$ implying that the exponent of the generalized
Debye spectrum is ${1/2}$, which contrast with the spectrum for
subdiffusion which is ${-1/4}$.

\section{Conclusion}
\label{Conclusion}

In this paper, we have introduced a general formalism offering a
multiscale description of transport processes in an intracellular
medium. By taking into account the forces exerted the
cystoskeleton and the protein motors in a generalized diffusion
equation, our model explains both enhanced diffusion and
subdiffusion observed in experiments in living cells. The
exponents associated to these behaviors are given in terms of the
parameters characterizing the system. These results could
be interpreted within a more general context in which the particle
is transported through the elastic medium undergoing a random path
through elastic scatters that induce a self-avoiding random walk
\cite{Adam-Rev}. This more general scenario permits to relate the
diffusion in the intracellular medium studied with other processes
in which self-avoiding random walks are an important ingredient.

Finally, we have proposed the density of states as a useful
quantity to describe the state of the cell. In the presence of
protein motors, this quantity behaves as the Debye spectrum with
an exponent larger than one. The model proposed may describe and
quantify the observed different forms of diffusion of particles in
a intracellular medium.

\vspace{0.5cm}
{\large Acknowledgments}
\vspace{0.25cm}

We acknowledge Prof. A. Gadomski for interesting discussions. This
work was supported by UNAM-DGAPA under the grant IN-102609 (ISH).


\clearpage

\begin{list}{}{\leftmargin 2cm \labelwidth 1.5cm \labelsep 0.5cm}

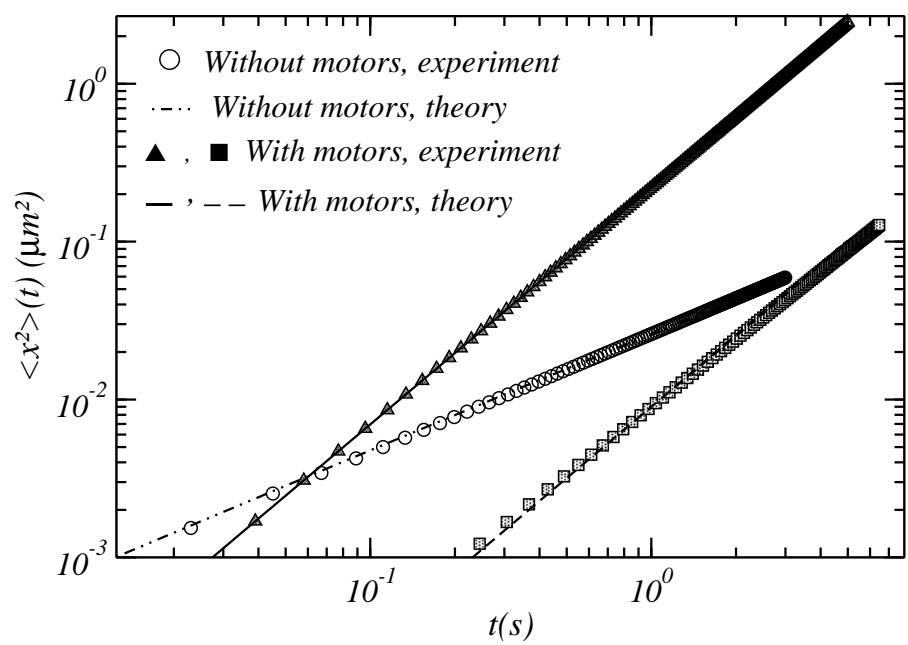
\item[\bf Fig. 1]  Mean square
displacements of microspheres engulfed in living eukaryotic cells.
Symbols represent data obtained with the fitting formula $\langle x^2\rangle = b t^{\alpha}$, obtained from experimental
measurements in Ref. \cite{caspi}. Circles correspond to the case of absence of motors with $b=0.026$ and $\alpha=0.75$. Triangles and squares correspond to the case of presence of motors with $b=0.22$, $\alpha=1.5$ and $b=0.009$,
$\alpha=1.42$, respectively.
The dot-dashed line (subdiffusion) is the theoretical result given
by Eq. (\ref{MSD2}) using (\ref{tau-Xi}) for the following values
of the parameters:
$D_0\simeq k_BT/m\beta_0\simeq 10^{-1}\mu m^2s^{-1}$, $\beta_0
\simeq 10^{7}s^{-1}$, $m \simeq 10^{-12}g$, $a\simeq 1\mu m$,
$\omega_0^2\simeq3.4\cdot10^{3}s^{-1}$, $B_2 \simeq 4.36$ and
$\tau_D \simeq 2.67s$. The dashed and solid lines show enhanced
diffusion obtained with Eq. (\ref{xx}) for
$\omega_1 \simeq 4.2\cdot10^{3}s^{-1}$ and $\omega_1 \simeq 3.4\cdot10^{3}s^{-1}$, respectively.

\end{list}

\clearpage

\begin{figure}[tbp]
{}
\par
\centering \mbox{\resizebox*{12cm}{!}{
\includegraphics{CaspiPRL-ExpTeor.eps}} }
\par
{\footnotesize {\ Figure 1. } \vspace{.8cm} }
\end{figure}

\end{document}